\documentclass[pdflatex,%
               sn-nature,%
               iicol,
               natbib=false
              ]{sn-jnl}

\usepackage[
 backend=biber,
  bibstyle=nature,
  citestyle=numeric-comp,
  autocite=superscript,
  sorting=none
]{biblatex}
\let\cite\autocite  

\DeclareRobustCommand{\citet}[1]{%
  \citeauthor{#1}~(\citeyear{#1})~\cite{#1}%
}
\usepackage[section]{placeins}
\usepackage{graphicx}%
\usepackage{multirow}%
\usepackage{amsmath,amssymb,amsfonts}%
\usepackage{amsthm}%
\usepackage{mathrsfs}%
\usepackage[title]{appendix}%
\usepackage{xcolor}%
\usepackage{stfloats}
\usepackage{tikz}
\usetikzlibrary{shapes.geometric, arrows.meta, positioning}
\usepackage{textcomp}%
\usepackage{manyfoot}%
\usepackage{booktabs}%
\usepackage{algorithm}%
\usepackage{algorithmicx}%
\usepackage{algpseudocode}%
\usepackage{listings}%
\usepackage[normalem]{ulem} 
\usepackage{xcolor}

\theoremstyle{thmstyleone}%
\theoremstyle{thmstyletwo}%

\theoremstyle{thmstylethree}%
\begin{document}

\title[Article Title]{Acoustic Classification of Maritime Vessels using Learnable Filterbanks}


\author*[1]{\fnm{Jonas} \sur{Elsborg}}\email{jels@dtu.dk}

\author[1]{\fnm{Tejs} \sur{Vegge}}

\author[1]{\fnm{Arghya} \sur{Bhowmik}}

\affil[1]{\orgdiv{Department of Energy Conversion and Storage}, \orgname{Technical University of Denmark}, \orgaddress{\street{Anker Engelunds Vej 301}, \city{Kongens Lyngby}, \postcode{2800}, \country{Denmark}}}


\abstract{Reliably monitoring and recognizing maritime vessels based on acoustic signatures is complicated by the variability of different recording scenarios. A robust classification framework must be able to generalize across diverse acoustic environments and variable source–sensor distances. To this end, we present a deep learning model with robust performance across different recording scenarios. Using a trainable spectral front-end and temporal feature encoder to learn a Gabor filterbank, the model can dynamically emphasize different frequency components. Trained on the VTUAD hydrophone recordings from the Strait of Georgia, our model, CATFISH, achieves a state-of-the-art 96.63\% test accuracy across varying source–sensor distances, surpassing the previous benchmark by over 12 percentage points. We present the model, justify our architectural choices, analyze the learned Gabor filters, and perform ablation studies on sensor data fusion and attention-based pooling.}

\maketitle
\section*{Introduction}\label{introduction}
Passive acoustic methods can detect vessels over long ranges because sound propagates efficiently in water, but the ocean environment’s intense ambient noise and multipath propagation cause significant signal attenuation and variability\cite{hildebrand2009anthropogenic, badiey2002temporal}. This problem is of practical interest as illegal fishing and vessel traffic in remote marine areas drive the need for autonomous acoustic monitoring systems\cite{domingos2022survey}. Yet real‐world recordings vary wildly: wind and waves mask tonal machinery signatures, frequency‐dependent attenuation and Doppler shifts distort spectra, and the same ship “sounds” different as its range to a hydrophone changes.  Fixed spectrogram‐based classifiers can achieve near‐perfect accuracy when training and test data share identical recording conditions, but performance collapses once data from multiple source–sensor distances or environments are mixed, representing a realistic use case scenario. Early passive-sonar work framed vessel recognition as a pattern-matching problem on hand-crafted descriptors such as LOFAR lines, MFCCs or gammatone coefficients, fed to Gaussian-mixture or SVM classifiers \cite{wu2014robust,li2017underwater,santos2016shipsear}. Larger public datasets and innovations in the field of machine learning (ML) led researchers to treat Mel or CQT spectrograms as images and apply mainstream image convolutional neural networks (CNNs) such as VGG\cite{choi2019acoustic}, ResNet\cite{domingos2022investigation}, DenseNet\cite{gao2020recognition} and MobileNet\cite{barros2022mobilenet} to ship-noise data. This boosted single-scenario accuracy on benchmark datasets such as ShipsEar\cite{santos2016shipsear} and DeepShip\cite{irfan2021deepship} into the mid-90\% range. Recently, end-to-end audio front-ends that learn Gabor- or Sinc-parameterised filters directly from the waveform, such as SincNet\cite{ravanelli2018interpretable}, LEAF\cite{zeghidour2021leaf}, and EfficientLEAF\cite{schluter2022efficientleaf}, have outperformed fixed filterbanks, and self-attention and transformer encoders such as Audio Spectrogram Transformer\cite{gong2021ast} and Swin Transformer\cite{xu2023self} have been shown to aid audio classification. However, in the Passive Underwater Acoustic Vessel Classification (PUAVC) task, robustness to real‐world variations is an open challenge, since cross-scenario evaluations in PUAVC still show double-digit accuracy drops when the source–sensor range or bathymetry changes — e.g. a fall from $\sim 94\%$ to 84\% on the VTUAD dataset when recordings from different distances are mixed \cite{domingos2022investigation}. A promising approach is to train learnable front-ends on data from multiple recording conditions, enabling the model to discover filterbanks that remain discriminative even when signals are severely attenuated or distorted by distance. Moreover, few existing models exploit environmental sensor metadata such as salinity and temperature, which could be useful since these variables are known to affect frequency-dependent transmission loss\cite{kuperman2014underwater,ferguson2017convolutional}. 
Inspired by these observations, we introduce the \textit{Classification Algorithm with Trainable Filterbanks for Identification of Ships} (CATFISH), an end‐to‐end framework that:  
\begin{itemize}
    \item Learns Gabor‐based filterbanks directly from raw waveforms,  
    \item Applies 2D attention pooling to emphasize propagation‐invariant spectral cues  
    \item Optionally fuses environmental variables to adapt to changing water conditions. In this work, these constitute Conductivity, Temperature, Depth, Salinity and Sound Velocity (jointly abbreviated as CTDSV). 
\end{itemize}
We evaluate CATFISH on the VTUAD multi‐scenario benchmark and demonstrate a $12$ percentage point gain over the benchmark fixed‐filter model, achieving $96.63$\% test set accuracy when trained on recordings at varied distances.
\section*{Results}\label{results}
\subsection*{Implementation and training}
\begin{figure}[!t]
    \centering
    \includegraphics[width=0.99\linewidth]{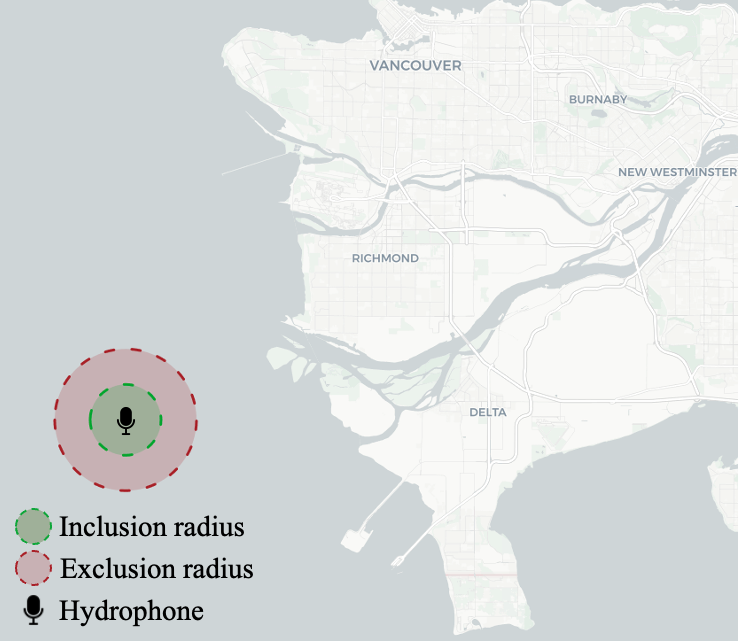}
    \caption{The VTUAD dataset contains hydrophone recordings from the Strait of Georgia off the coast of Vancouver, grouped in three scenarios (see table below). The methodology is such that recordings were added whenever a single vessel was inside the inclusion zone, and no other vessel was within the exclusion zone. For the \textit{Background} class, no ships are inside the exclusion radius.}
    \label{fig:inc_exc}

    \vspace{1em}

    \begin{tabular}{lcc}
    \toprule
    Scenario & Inclusion radius & Exclusion radius \\
    \midrule
    S1 & 2 km & 4 km \\
    S2 & 3 km & 5 km \\
    S3 & 4 km & 6 km \\
    \bottomrule
  \end{tabular}
\end{figure}

\begin{figure}
  \centering
  \begin{tikzpicture}[
      font=\footnotesize,
      mod/.style={
        draw,
        rounded corners=2pt,
        minimum height=9mm,
        align=center,
        inner sep=3pt,
      },
      frontend/.style={mod, fill=blue!10},
      backbone/.style={mod, fill=green!10},
      fusion/.style={mod, fill=orange!12},
      head/.style={mod, fill=purple!10},
      meta/.style={mod, dashed, fill=orange!20},
      arr/.style={-{Stealth[length=3pt]}, thick},
      node distance=4mm and 5mm
    ]
    \node[frontend] (raw)            {Raw\\Waveform};
    \node[frontend, below=of raw] (gabor) {96× Learnable\\Gabor Filters};
    \node[frontend, below=of gabor] (gauss) {Gaussian\\Pooling};
    \node[frontend, below=of gauss] (pcen)  {Log-Compression\\+ TBN};

    \node[backbone, below=8mm of pcen] (mbconv) {EfficientNet-B0\\\textbf{MBConv × 7}};

    \node[backbone, below=of mbconv]  (attn) {2-D Attention\\Pooling};

    \node[meta, right=10mm of attn, yshift=6mm] (ctd) {CTDSV Data};

    \node[fusion, below=of attn] (concat) {Concatenate};
    \node[head, below=of concat] (fc) {FC + Softmax Classification Head};

    \foreach \s/\t in {raw/gabor, gabor/gauss, gauss/pcen,
                       pcen/mbconv, mbconv/attn, attn/concat, concat/fc}
        \draw[arr] (\s) -- (\t);

    \draw[arr] (ctd.west) |- (concat.east);

    \node[left=2mm of gabor.west, anchor=east, font=\scriptsize\bfseries, yshift=3mm] {Frontend};
    \node[left=2mm of mbconv.west,anchor=east, font=\scriptsize\bfseries, yshift=3mm] {Backbone};
    \node[left=2mm of attn.west,  anchor=east, font=\scriptsize\bfseries, yshift=3mm] {Pooling};
    \node[left=2mm of concat.west,anchor=east, font=\scriptsize\bfseries, yshift=3mm] {Fusion};
    \node[left=2mm of fc.west,    anchor=east, font=\scriptsize\bfseries, yshift=3mm] {Head};
  \end{tikzpicture}
  \caption{End-to-end architecture of the CATFISH model. Audio passes through a learnable Gabor-based frontend and EfficientNet-B0 backbone, followed by 2D attention pooling. Environmental metadata (CTDSV) can be fused before the final fully-connected classification head.}
  \label{fig:catfish_arch}
\end{figure}
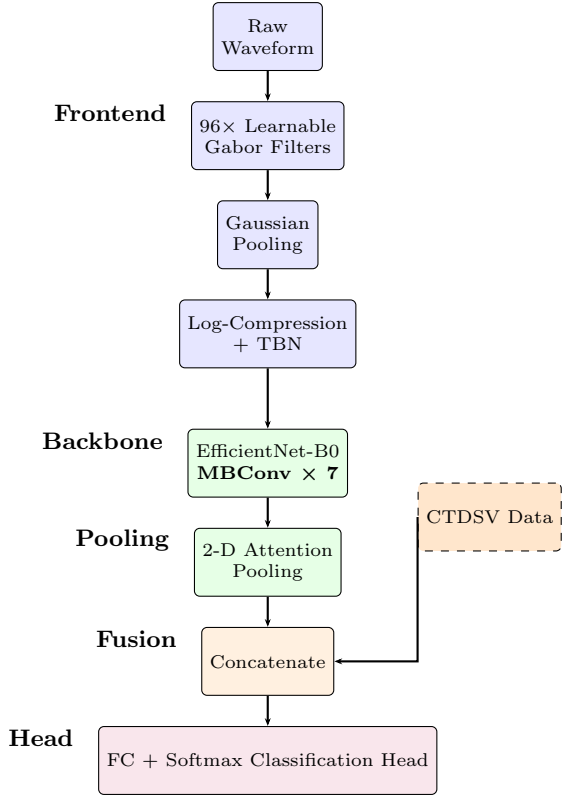

Our experiments employ the VTUAD benchmark, which comprises 1-second hydrophone clips from the Strait of Georgia off the coast of Vancouver and Richmond. The audio in each clip is annotated either as one of four vessel classes (\textit{Tug, Tanker, Cargo, Passengership}) if a vessel is present, or as a \textit{Background} class if the recording consists of underwater ambient noise. The data was collected under three distance‐based scenarios to test robustness across varying source–sensor ranges. Figure~\ref{fig:inc_exc} illustrates the definition of a scenario. 
We train all variations of the CATFISH model for 40 epochs on a single NVIDIA RTX 3090-24GB GPU. The training time is no more than 8 hours for a single model trained on all scenarios. The main CATFISH model uses both attention pooling and injects the CTDSV sensor data into the final classification head as described above. The architecture is depicted in Figure \ref{fig:catfish_arch}, and is trained against a multi-class cross-entropy loss. In Table~\ref{tab:comparison}, we report the test set accuracies obtained when training the model on each scenario separately as well as jointly on all scenarios. This methodology follows the original benchmark from~\citet{domingos2022investigation}. 
As shown in the table, CATFISH outperforms all but one prior model on the individual scenarios by a significant margin. The exception is scenario 1, where~\citet{li2024enhancing}
achieved 98.15\% test set accuracy, while CATFISH reaches 97.55\%. More importantly, CATFISH reaches a test set accuracy of 96.63\% on the multi-scenario test set. Thus, CATFISH outperforms previous state-of-the-art on the VTUAD dataset and achieves a 12 percentage point (pp) gain compared to the benchmark of 84.13\%~\cite{domingos2022investigation}.
\setlength{\tabcolsep}{3.5pt} 
\begin{table}[!t]
\centering
\caption{Comparison of the main CATFISH model's test accuracy\% with the original benchmark from~\citet{domingos2022investigation}, as well as models from subsequent publications using the VTUAD data~\cite{li2024enhancing,nathala2024vessel}. For all numbers from previous publications, a model was trained only on data from the corresponding scenario. For CATFISH, we report two accuracies for each scenario; CATFISH (single) refers to the accuracy for models trained only on single scenarios, while CATFISH (combined) refers to the accuracy of the combined-scenario model when evaluated on data from single scenarios. }
\begin{tabular}{lcccc}
\noalign{\vskip 2mm}
\hline
\noalign{\vskip 2mm}
Source & S1 & S2 & S3 & All \\
\noalign{\vskip 2mm}
\hline
\noalign{\vskip 2mm}
\citet{domingos2022investigation} & 94.95 & 94.45 & 93.11 & 84.13  \\
\citet{li2024enhancing} & \textbf{98.15} & - & - & -  \\
\citet{nathala2024vessel} & - & - & 93.53 & -  \\
CATFISH (single) & 97.55 & \textbf{97.46} & 95.03 & -  \\
CATFISH (combined) & 96.01 & \textbf{97.46} & \textbf{95.98} & \textbf{96.63}  \\
\hline
\end{tabular}
\label{tab:comparison}
\end{table}
\subsection*{Attention pooling and environmental variables}
To separate the effects of the CTDSV data and the attention pooling, we ablated these two components. To ablate the attention pooling, we trained models that instead use the default global max pooling from EfficientLEAF~\cite{schluter2022efficientleaf}. To ablate the effect of the CTDSV data we removed the CTDSV classification head. The results are shown in Table \ref{tab:ablation}, and demonstrate that the addition of the learnable frontend without attention or CTDSV data still yields an improvement on the combined scenario, with a test set accuracy of 91.59\% (7.46 pp higher than the benchmark). Furthermore, the addition of either CTDSV or attention brings the accuracy to over 96\%. Including both attention and CTDSV does not improve accuracy significantly (96.63\% vs 96.23\% for attention only, and 96.32\% for CTDSV only). This indicates that either of these two mechanisms can inject the information necessary to discern recordings from varying distances.
\begin{table}[ht!]
\centering
\caption{Ablation study of the attention pooling mechanism from CATFISH versus the default max pooling from the LEAF models, as well as the inclusion of the CTDSV data. All numbers for single scenarios refer to the accuracy obtained when the model is trained exclusively on data from that scenario.}
\begin{tabular}{lcccc}
\noalign{\vskip 2mm}
\hline
\noalign{\vskip 2mm}
CATFISH Model & S1  & S2 & S3 & All \\
\hline
\noalign{\vskip 2mm}
\multicolumn{5}{l}{Attention pooling} \\
\hline
\noalign{\vskip 2mm}
96 filters & 94.78 & 95.02 & 92.35 & 96.23  \\
96 filters + CTDSV & \textbf{97.55} & \textbf{97.46} & 95.03 & \textbf{96.63}  \\
\hline
\noalign{\vskip 2mm}
\multicolumn{5}{l}{Max pooling} \\
\hline
\noalign{\vskip 2mm}
96 filters & 96.11 & 93.27 & 93.08 & 91.59  \\
96 filters + CTDSV & 97.18 & 96.11 & \textbf{96.10} & 96.32  \\
\hline
\end{tabular}
\label{tab:ablation}
\end{table}
\subsection*{Filter activation variability with class and scenario}
\begin{figure}[t!]
    \centering
    \includegraphics[width=0.97\linewidth]{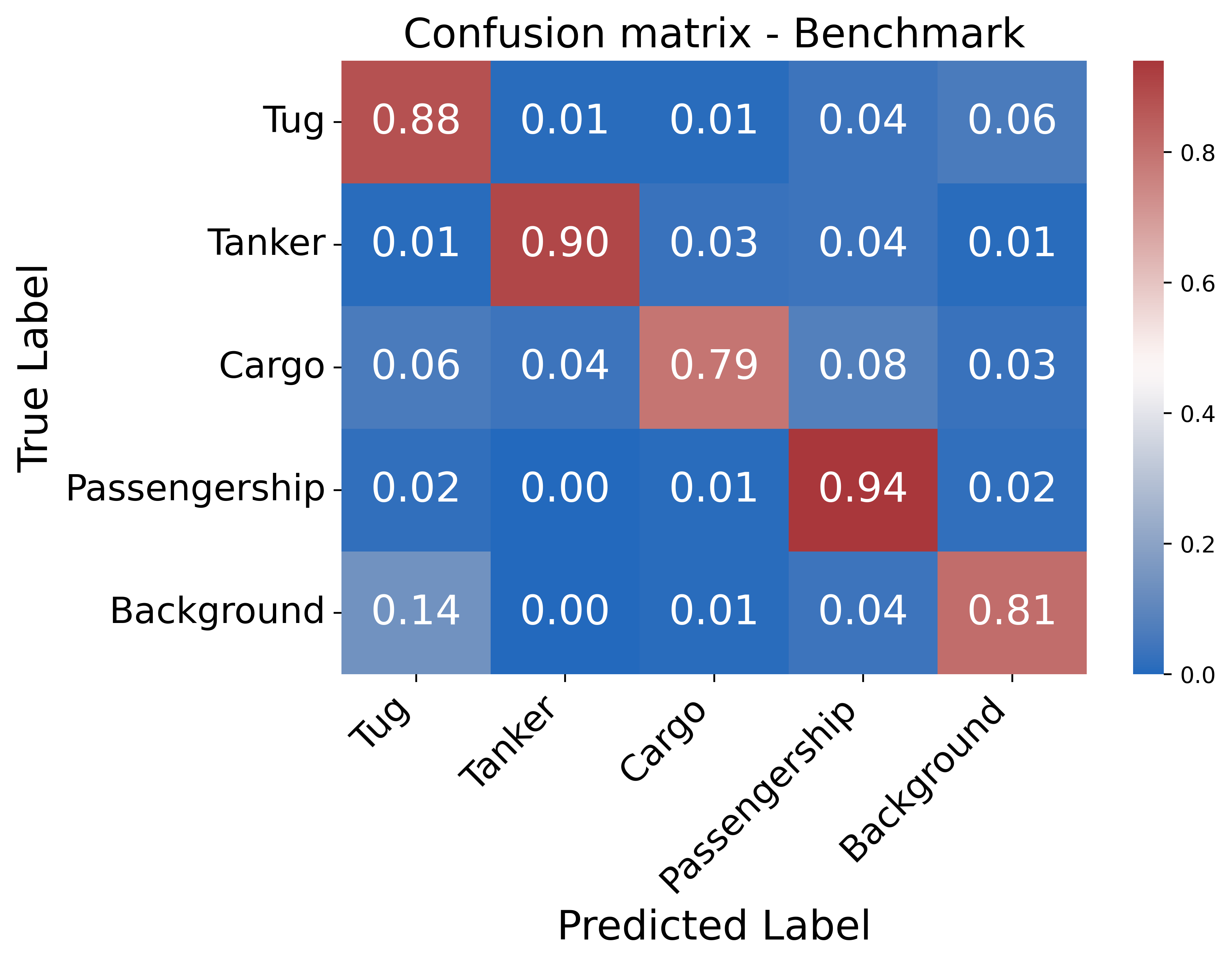}\\[1.2em]
    \includegraphics[width=0.97\linewidth]{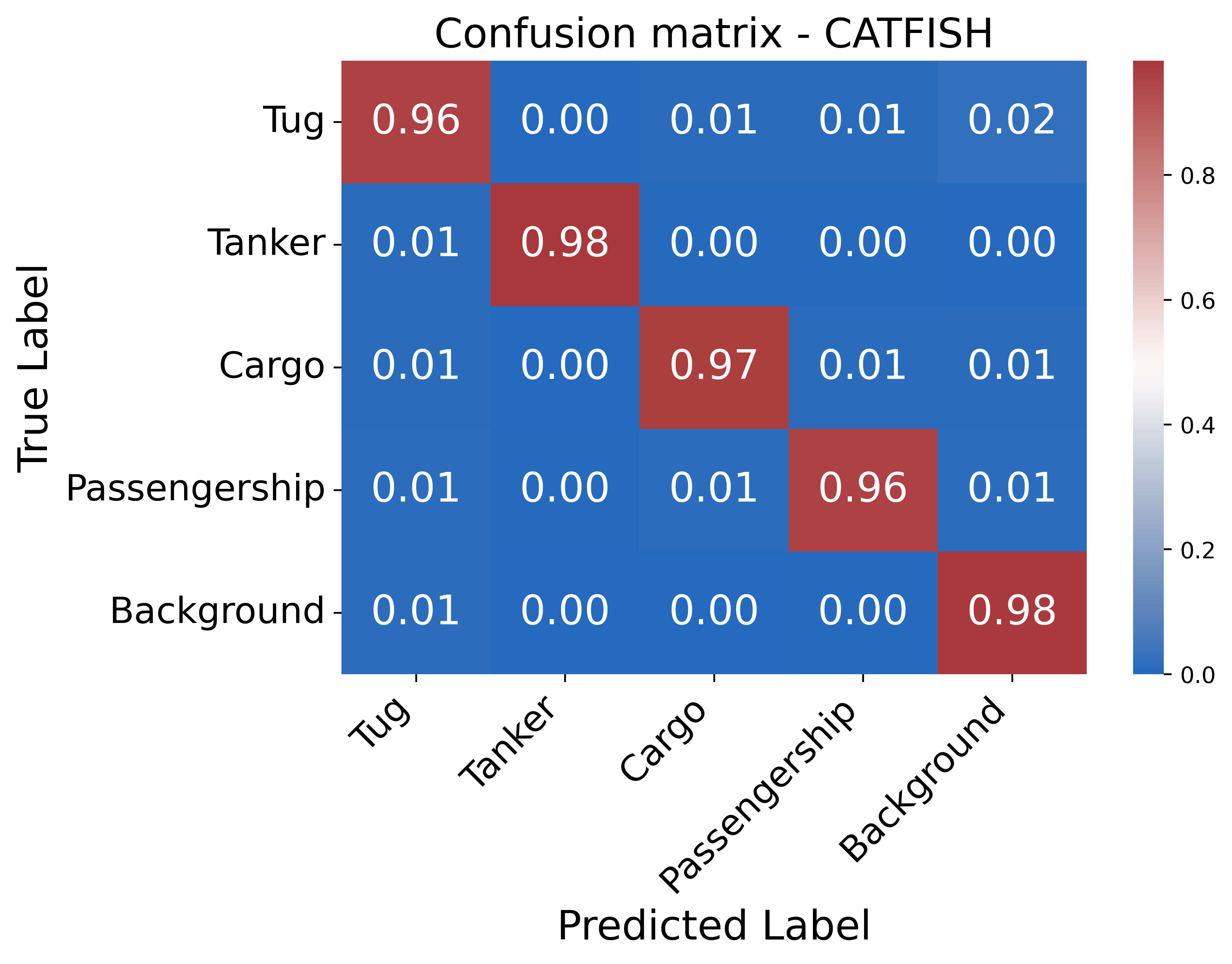}
    \caption{\textbf{Top}: Test set confusion matrix from the original benchmark on the multi-scenario task, reproduced from \citet{domingos2022investigation}. \textbf{Bottom}: Test set confusion matrix for CATFISH on the multi-scenario task. }
    \label{fig:conf_matrix}
\end{figure}

\begin{figure}[!t]
    \centering
    \includegraphics[width=0.99\linewidth]{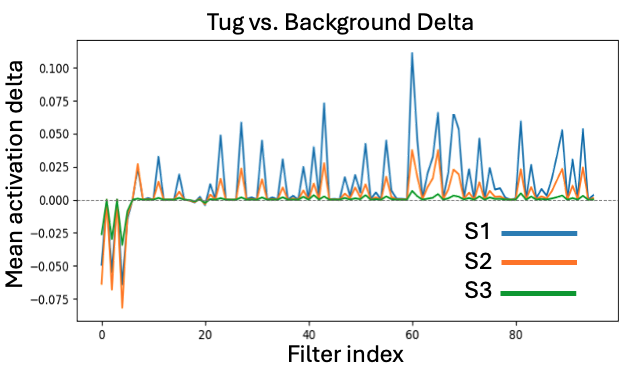}
    \caption{Mean activation delta between tug and background for each learned filter across all three recording scenarios.}
    \label{fig:freq_response}
\end{figure}

\begin{figure}[ht!]
    \centering
    \includegraphics[width=0.99\linewidth]{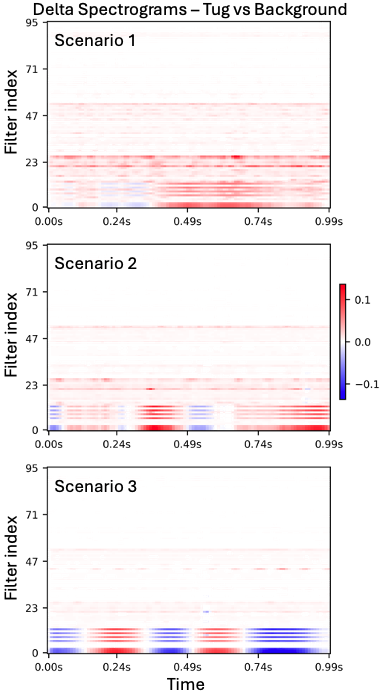}
    \caption{$\Delta$-spectrograms (Tug-Background) over filter index (0-95) and time (0-1s) for the three scenarios, averaged over all test clips. The range of active filters contracts from $\approx 0-60$ in Scenario 1 to $\approx 0-45$ in Scenario 2 and finally to filters $\approx 0-10$ in Scenario 3, which are consistently strong in all scenarios.}
    \label{fig:freq_diff_2d}
\end{figure}

Classes with similar frequency ranges are the most frequently confused classes in the original benchmark. In particular, the \textit{Tug} and \textit{Background} classes have large errors, with 14\% of background waveforms predicted as tugs and 6\% of tug waveforms predicted as background noise~\cite{domingos2022investigation} (See Figure~\ref{fig:conf_matrix} comparing the confusion matrix for the original multi-scenario benchmark versus that of CATFISH). For CATFISH, the corresponding numbers are 2.2\% and 1.9\%.
To understand why the learned Gabor filterbank achieves this vast improvement, we investigate how the filters respond to different vessel signatures. We first pass each training waveform $x$ through the filterbank alone, producing an activation tensor  
\begin{equation}
    A^{(k)}_{c,t} \quad\bigl(c=1,\dots,C,\;t=1,\dots,T\bigr),
\end{equation}
where $c$ indexes the $C$ filters (increasing $c$ → higher center frequency) and $t$ indexes time frames. We focus on the \textit{Tug} and \textit{Background} classes (Bg), and average over time to obtain class‐conditional mean activations
\begin{equation}
    \overline{a}_c^{\,\text{Tug}} = \frac{1}{N_{\text{T}}\,T}\sum_{n\in\text{Tug}}\sum_{t}A^{(n)}_{c,t},
\end{equation}
\begin{equation}
    \overline{a}_c^{\,\text{Bg}} = \frac{1}{N_{\text{B}}\,T}\sum_{n\in\text{Bg}}\sum_{t}A^{(n)}_{c,t},
\end{equation}
and their difference  
\begin{equation}
    \Delta_c = \overline{a}_c^{\,\text{Tug}} - \overline{a}_c^{\,\text{Bg}}.
\end{equation}
Figure~\ref{fig:freq_response} plots $\Delta_c$ as a function of filter index $c$, providing a compact ranking of each filter’s overall discriminative power. 
It demonstrates that the learned filters provide clear deltas, even for Scenario 3 where the discriminative signal is weakest. 
To visualize this in more detail, Figure~\ref{fig:freq_diff_2d} shows the full two‐dimensional $\Delta$‐spectrogram  
\begin{equation}
    \Delta_{c,t} = \bigl\langle A_{c,t}\bigr\rangle_{\text{Tug}} - \bigl\langle A_{c,t}\bigr\rangle_{\text{Bg}},
\end{equation}
displaying the element‐wise difference of mean activations across time $t$ and filter channels $c$, averaged over clips.
In all three inclusion/exclusion settings, discriminative power remains concentrated in the lowest‐index filters, but the breadth of active channels shrinks with distance. In Scenario 1 (top), red and blue regions span a wide band (indices $\approx$0–60), with strong tug signatures in mid‐bands ($\approx$10–30). In Scenario 2 (middle), activity narrows to indices $\approx$0–45, preserving the mid‐band tug peak but reducing high‐index responses. In Scenario 3 (bottom), only the very lowest filters (0–10) retain reliable Tug vs Background contrast, as more distant vessels attenuate higher‐frequency cues. Thus, while filters 0–10 are robust across all scenarios, the model adaptively contracts its focus from a broad low‐frequency range toward just those ultra‐low‐index channels that remain informative at greater distances. These results demonstrate that end‐to‐end Gabor‐filter learning can recover stable spectral cues even when ships move across very different source–sensor distances, resulting in CATFISH setting a new state‐of‐the‐art within this area, as shown in Table \ref{tab:comparison}.
\section*{Methods}
\subsection*{Dataset and challenge}
Several datasets have been used for the PUAVC task. The largest corpus of research has focused on the ShipsEar~\cite{santos2016shipsear} dataset. ShipsEar consists of 90 sound recordings made in 2012 and 2013 off the Spanish Atlantic coast. Each recording is labeled according to one of five vessel classes (including a “no-vessel” background class), making ShipsEar a well-defined, yet relatively straightforward benchmark where test set accuracies of over 99\% has been achieved~\cite{wei2025underwater}. DeepShip~\cite{irfan2021deepship} and VTUAD (Vessel Type Underwater Acoustic Data)~\cite{domingos2022investigation} are two similar datasets that are both based on recordings from hydrophones deployed by Ocean Networks Canada (ONC). DeepShip consists of over 47 hours of recordings from four different vessel types, as well as background recordings. VTUAD consists of roughly 49 hours of recordings, split into 1-second clips. However, DeepShip and VTUAD differ in two important ways:
\begin{itemize}
    \item DeepShip only includes recordings where a single vessel is within 2 km of the hydrophone. In contrast, VTUAD includes data from three different scenarios defined by inclusion and exclusion radii (see Figure~\ref{fig:inc_exc}). 
    \item DeepShip consists of four vessel classes (Tug, Tanker, Cargo, Passengership) from the same source. This data set also includes a background class, but crucially, data for this class were added from a separate source. In contrast, VTUAD has an additional fifth background noise class from the same source.
\end{itemize}
The first difference is emphasized by the creators of VTUAD as a way to enable investigation of the effects of the expected lowering of the signal-to-noise ratio (SNR) of recordings where the vessel is further away. Additionally, including a background noise class from the same source is essential for robustness testing, as it is not clear to what degree the noise is location- and source-dependent. Drawing the background data from a different corpus than the vessel recordings creates a domain shift between the vessel and background classes\cite{gourishetti2022potentials}. Consequently, the background recordings may not have the same noise floor, reverberation, and sensor characteristics, and the network can learn dataset quirks rather than focusing on the actual acoustic signatures. Thereby the model can simply learn to tell apart the two datasets rather than genuinely detecting the presence or absence of a ship, so while recent work has surpassed 99 \% test accuracy\cite{li2022underwater} on the DeepShip dataset, these results are less broadly applicable because of the weaknesses above. The single scenario of DeepShip has an inclusion radius of 2 kilometers, which can safely be assumed to have a better signal-to-noise ratio than scenarios with more distant vessels (as also seen in the VTUAD results in Table \ref{tab:comparison}). For this reason and because of the domain-shifted background class, it is not possible to conclude whether the best models on DeepShip would perform well on a more realistic dataset with recordings from various distances and a more challenging background class. This is corroborated by the original VTUAD paper, which shows that even a carefully crafted combination of preprocessing filter, network architecture, and optimizer suffers from a performance drop when trained on data from all three scenarios~\cite{domingos2022investigation} versus only a single scenario. Indeed, the single-scenario accuracies can be as high as 97\%, while the accuracy on the multi-scenario data was around 84\%, even with the best possible fixed-filter model. In other words, the model performance drops severely when recordings do not originate from roughly the same distance. As demonstrated in the paper, it is the background class which confounds the model when scenarios are mixed, indicating that the difference in SNR between the scenarios cannot be resolved by a fixed filter. Naturally, any real usecase of PUAVC must be able to handle the constant presence of background noise while being robust to the reality that vessels approach and move at different distances from the recording device. Thus, the combined scenario of the VTUAD dataset is the most challenging task from the canonical PUAVC datasets, as well as the most useful and realistic benchmark. Accordingly, it is this dataset we used as the basis for the model presented in this work. 
\subsection*{Model}
We propose a model that is more robust to varying scenarios by constructing a network that learns its front-end filters, temporal encoding, and metadata fusion jointly from raw waveforms. Thus, rather than using fixed filters, we employ a learnable filterbank initialized as Gabor filters, allowing the model to tune frequency bands to the data. This approach is inspired by prior neural filterbanks that learn task-specific audio filters~\cite{ravanelli2018interpretable}, and our model is based on the EfficientLEAF~\cite{schluter2022efficientleaf} model. As in the original LEAF implementation~\cite{zeghidour2021leaf}, this is a supervised classification problem that jointly learns the classification model parameters $\theta$ and the frontend parameters $\psi$:
\begin{equation}
\theta^*, \psi^*=\underset{\theta, \psi}{\arg \min } E_{(x, y) \in \mathcal{D}} \mathcal{L}\left(g_\theta\left(\mathcal{F}_\psi(x)\right), y\right),
\end{equation}
where $\mathcal{F}_\psi(x)$ is the frontend representation (a learnable filterbank) of the raw waveform $x$, and y is the label of sample $(x,y)$ from the dataset $\mathcal{D}$. 
We refer to the original publications~\cite{zeghidour2021leaf,schluter2022efficientleaf} for the full details on the learnable Gabor filters of LEAF and EfficientLEAF. Briefly, the full set of learnable frontend parameters is
\begin{equation}
    \psi = \{\mu_{k}, \sigma_{k}, \rho_{k}, \alpha_{k}, \gamma_{k}, \beta_{k}\}, k=1,...,K
\end{equation}
where $(\mu_{k}, \sigma_{k})$ is the Gabor center frequency and (inverse) bandwidth for each of K filters, and $\rho_k$ is a trainable Gaussian-pooling window. In EfficientLEAF, the Per-Channel Energy Normalization (PCEN) layer from the original LEAF code is replaced by a fully parallel compression block that itself has only learnable parameters: first a trainable per-band log‐gain $a_{k}$ in:
\begin{equation}
y_k[t]=\log \left(1+10^{a_k} x_k[t]\right),
\end{equation}
and then a Temporal BatchNorm (TBN) with per-band affine weights $\gamma_k,\beta_k$. 
The model $g_\theta(\cdot)$ processes the frontend $\mathcal{F}_\psi(x)$ through an EfficientNet-B0~\cite{tan2019efficientnet}
embedding backbone. In both LEAF and EfficientLEAF, this embedding undergoes global max pooling and a linear classification layer. To capture salient time–frequency patterns, our model incorporates an attention pooling mechanism\cite{vaswani2017attention}. This design follows recent trends in audio classification, where self-attention has been shown to improve performance on spectrogram inputs. For example, \citet{gong2021ast} introduced Audio Spectrogram Transformer, which applies a ViT/Transformer with self-attention to audio spectrograms, achieving state-of-the-art results on the AudioSet dataset. Furthermore, CATFISH can optionally use a set of normalized environmental variables which are canonically abbreviated as CTD (Conductivity, Temperature, Depth). In the VTUAD dataset, salinity and sound velocity data are also available, and thus we denote the inclusion of all five variables by CTDSV. If included in the model, they are processed by a small feed-forward branch and concatenated with the audio embedding before classification.
\section*{Discussion}
Contrary to the mixed results of EfficientLEAF on general audio tasks~\cite{schluter2022efficientleaf}, our underwater experiments show that the learnable frontend accounts far better for the frequency‐dependent attenuation and noise masking that complicate the robustness to distant vessel recordings~\cite{domingos2022investigation}. The ablations reveal that simply replacing a fixed Mel filterbank with trainable filters yields most of the 12 pp accuracy gain; adding either 2D attention pooling or CTDSV metadata contributes an additional ~5 pp by either focusing on invariant tonal patterns or providing environmental context for distorted bands. In practice, attention pooling may be preferred when environmental sensors are unavailable, while CTDSV fusion offers a lightweight side‐channel for water‐condition adaptation. However, the extra parameters (around 4.6 M for the filterbank plus attention layers) and training complexity require GPU‐oriented pipelines for model updates, and quantization or pruning may be necessary for on‐device inference. Moreover, safety‐critical applications (e.g. port security) demand calibrated confidence estimates. Bayesian neural networks or deep ensembles could supply reliable uncertainty bounds, enabling human review of low‐confidence detections. Finally, while VTUAD covers three range scenarios in one geographic region, true field deployments must handle seasonal thermocline shifts, varying seabed composition, and entirely new vessel classes. Future work should explore unsupervised domain adaptation (e.g.\ self‐supervised pretraining on unlabelled hydrophone streams), continual learning for emerging vessel types, and the interpretability of learned filters in consultation with marine acoustics experts. Integrating CATFISH with AIS geolocation data and multi‐sensor fusion will be key to building robust, autonomous underwater surveillance networks that generalize beyond the Strait of Georgia.

\section*{Data availability}
The VTUAD dataset used in this work is available from IEEE DataPort (DOI: 10.21227/msg0-ag12). 

\section*{Code availability}
The computer code necessary to train and evaluate the CATFISH model is freely available on GitHub\footnote{\url{https://github.com/Jotels/CATFISH}}.


\printbibliography
\section*{Acknowledgements}
Authors acknowledge financial support from the Danish National Research Foundation with the Pioneer Center for Accelerating P2X Materials Discovery (CAPeX) (Grant No. P3).

\section*{Author contributions}
J.E., T.V., A.B. worked on the conceptualisation, J.E. collected, analysed  and visualized the data.
T.V., A.B. acquired funding and resources. J.E. wrote the original
manuscript draft. All authors reviewed the manuscript.
\section*{Competing interests}
The authors declare no competing interests.

\end{document}